\documentclass[prb,aps,amssymb, amsmath,preprint,showpacs]{revtex4}

\usepackage{graphicx}
\usepackage[OT1]{fontenc}
\usepackage{float}
\usepackage{epsfig}

\begin{document}

\title{Transport and noise in resonant tunneling diode
using self-consistent Green function calculation}

\author{V. Nam Do and P. Dollfus  }
\affiliation{Institut d'Electronique Fondamentale  \\
B$\hat{a}$timent 220 - Universit\'{e} Paris Sud, 91405 Orsay
cedex, France}

\author{V.Lien Nguyen}
\affiliation{Theoretical Dept., Institute of Physics, VAST \\
P.O. Box 429  Bo Ho, Hanoi 10000,  Vietnam }

\begin{abstract}
The fully self-consistent non-equilibrium Green functions (NEGFs)
approach to the quantum transport is developed for the
investigation of one-dimensional nano-scale devices. Numerical
calculations performed for resonant tunneling diodes (RTDs) of
different designs and at different temperatures show reasonable
results for the potential and electron density profiles, as well
as for the transmission coefficient and the current-voltage
characteristics. The resonant behavior is discussed in detail with
respect to the quantum-well width, the barrier thickness, and the
temperature. It is also shown that within the framework of
approach used the current noise spectral density can be
straightforwardly calculated for both the coherent and the
sequential tunneling models. In qualitative agreement with
experiments, obtained results highlight the role of charge
interaction which causes a fluctuation of density of states in the
well and therefore a noise enhancement in the negative
differential conductance region.
\end{abstract}


\maketitle

\section {Introduction }

In nano-scale semiconductor devices the quantum effects become
increasingly important and may dominate the transport phenomena.
Whatever the system under study the traditional semiclassical
Boltzmann equation is no longer an adequate approximation. A
rigorous quantum-mechanical approach is now necessary not only to
study devices intrinsically based on quantum phenomena (e.g.
resonant structures), but also to describe more conventional
devices as nano scaled field effect transistors where quantum
effects cannot be neglected anymore. However, the quantum
transport theories are more difficult to implement within a
general framework. Among the different quantum formalisms
developed during the last decades,\cite{ferry} the Non-Equilibrium
Green Functions can be regarded as one of the most powerful due to
its transparent physical route. It provides a general approach to
describing quantum transport in the presence of scattering,
including the contact and/or gate couplings. A very pedagogical
review of the subject can be found in,\cite{datta} where the
procedure of the self-consistent solution of the Green function
and the Poisson equation is properly described. As stated by
Christen and B\"{u}ttiker,\cite{christ} the self-consistent
treatment of the Coulomb potential is necessary to ensure
gauge-invariant current-voltage (I-V) characteristics in
non-linear transport.

However, though the formalism is clear, the NEGF-calculation of
transport quantities is often technically complicated. Except a
few studies realizing the full two-dimensional
calculation,\cite{sviz} it is mostly developed for one-dimensional
(1D) transport problems possibly coupled with the two- or
three-dimensional description of
electrostatics.\cite{venu,wang,besc} The resonant tunneling diode
is a typical 1D quantum structure which has been extensively
studied both theoretically and experimentally. Although the basic
structure of this device is simple, the development of a fully
self-consistent NEGF-calculation likely to give reasonable I-V
characteristics is still a subject of many efforts.
\cite{potz,lake92,lake97} Alternatively, the Wigner function
formalism \cite{kluk,shif,nedj} or the self-consistent solution of
the Schr\"{o}dinger and Poisson equations \cite{pina} have been
also used to calculate transport properties of RTDs.

The aim of this work is to present results of the fully
self-consistent NEGF-calculation of typical electronic and
transport quantities such as potential and electron density
profile, transmission coefficient, I-V characteristics, and
current noise spectral density in double barrier resonant
tunneling structures. The Green functions are calculated exactly
within the framework of the tight-binding model and
self-consistently coupled with the Poisson equation. The
calculation procedure though standard is formulated in a simple
form that can be readily applied to any 1D nano-scale devices and
can be easily extended to more complicated structures. Compared to
other works in the literature, we obtain a good description of
resonant behavior for RTDs of different designs and at different
temperatures. Although the NEGF-method has been used to calculate
the noise in various structures,\cite{levy,ding} to our best
knowledge, the present work is the first attempt to use this
formalism to show both sub-poissonian and super-poissonian shot
noise in RTDs, which is now the subject of great attention, both
theoretically \cite{martin,ianna97,ianna98,blant99,blant20,oriols}
and experimentally.\cite{kuzn,song,tsu,nau,new}

The paper is organized as follows. Section II describes the model,
formulates the problem and the calculation method. In section III
the numerical results of the potential and electron density
profiles, the transmission coefficient, and the I-V
characteristics are presented and discussed in comparison with
those reported in the literature. Section IV is specially devoted
to the noise calculation. Throughout the work, an attention is
properly paid to the practical manner of calculation technique.

\section{ Method and Formulation }
Let us consider a double barrier resonant tunneling device
connected to two infinite contacts, the left (L) and the right (R)
(Fig. 1). We assume that each contact is characterized by an
equilibrium Fermi function, $f_{L(R)} = 1/ [\exp (E - \mu_{L(R)}
)/k_B T + 1]$, with a chemical potential, $\mu_{L(R)}$. The
$x$-direction is chosen to be perpendicular to the barriers. The
$(y-z)$-cross-section is assumed to be large so that the potential
can be considered translation-invariant in this plane. The device
is then described by the standard single-band effective mass
Hamiltonian:
\begin{equation}
H = H_{||} + H_\perp  = [ E_c - \frac{\hbar^2}{2m_{\parallel}}
\frac{d^2}{dx^2} + U(x) ] - \frac{\hbar^2}{2m_\perp}
\Delta_\perp^2 \ , \label{1}
\end{equation}
which is coupled to two contacts. In Eq. (1), $E_c$ includes the
bottom of conduction band and the conduction-band discontinuities,
$m_{\parallel} (m_\perp )$ is the longitudinal (transverse)
electron effective mass, and $U(x)$ is the total electrostatic
potential. Within the framework of the single particle model, the
potential $U(x)$ can be determined from the Poisson equation:
\begin{equation}
\frac{d}{dx} ( \epsilon \frac{dU(x)}{dx} ) = \frac{e^2}{\epsilon_0
} [ n_d (x)- n_e (x) ], \label{2}
\end{equation}
where $e$ is the electron charge, $\epsilon_0$ is the dielectric
constant of vacuum, $\epsilon$ is the relative dielectric constant
which may vary along the $x$-direction, according to the material,
and $n_e$ and $n_d$, respectively, are the electron and donor
density, which are assumed to be independent of $y$- and
$z$-coordinates. As usual, we consider the case of fully ionized
donors.

In order to solve the Hamiltonian (1) and Eq. (2)
self-consistently with respect to the potential $U$ and the
electron density $n_e$, taking into account the contact couplings,
and further, to calculate the quantities of interest such as the
electron density profile and I-V characteristics of the device, it
is convenient to use the NEGF method. For the numerical procedure
the system under study is spatially meshed and all quantities of
interest are computed at the grid sites. For a grid spacing $a$
along the $x$-direction, since the Hamiltonian (1) is already
variable-separated, we can write it in the matrix form:
\begin{equation}
[H]_{i, {\bf k}, i', {\bf k'}} = -\left[t_x \delta_{i+1,i'} -
(2t_x + U_i + E_{ci} + \varepsilon ({\bf k}) ) \delta_{i,i'} + \
t_x \delta_{i-1,i'}\right] \ \delta_{{\bf k},{\bf k'}} ,\label{3}
\end{equation}
where $i (i')$ indicates grid sites, numbered from left $(1)$ to
right $(N)$, $t_x = \hbar^2/2m_{\parallel} a^2$ is the coupling
energy between adjacent sites along the $x$-direction (in the
nearest neighbor tight-binding scheme), and $\varepsilon ({\bf k})
= \hbar^2 k^2/ 2m_\perp$ with ${\bf k}$ being the wave vector in
the $(y-z)$-plane. The Hamiltonian matrix (3) is written in the
$\{\mid i> \bigotimes\mid {\bf k}> \}$-basis.\cite{datta}

Once we have a matrix representation of the Hamiltonian operator
it may seem straightforward to get the retarded Green function
(GF) by inverting the matrix, $[(E - i\eta )I - H ]^{-1}$, where
$I$ is the unit matrix and $\eta$ is an infinitesimal positive
real quantity, $\eta \rightarrow 0^+$. In practical calculations,
however, avoiding to work with infinite contacts, the GFs are
calculated only in the device domain and the contact couplings are
introduced as the 'self-energy' matrices $\Sigma^r$. Thus, for the
device under study, taking into account the couplings with two
contacts, L and R, the retarded GF can be determined as
\begin{equation}
G^r = [ (E - i\eta )I - H - \Sigma^r_L - \Sigma^r_R ]^{-1}
,\label{4}
\end{equation}
where $\Sigma^r_{L(R)}$ is the retarded self-energy matrix
describing the coupling between the device and the left (right)
contact. Within the framework of the 1D nearest-neighbor tight
binding model, these self-energy matrices can be calculated
exactly:
\begin{equation}
[\Sigma^r_{L(R)} ]_{i,{\bf k},i',{\bf k'}} = t_x^2 [g_{L(R)}
(E,{\bf k})]_{i,i'} \delta_{{\bf k},{\bf k'}} \delta_{i,i'}
\delta_{i,1(N)}, \label{5}
\end{equation}
where $g_{L(R)} (E,{\bf k})$ is a solution of the equation:
\begin{equation}
t_x^2 [g_{L(R)}]^2 - \alpha_{L(R)}(E, {\bf k}) g_{L(R)} + 1 =
0,\label{6}
\end{equation}
with the convenient sign for the root. In this equation
$\alpha_{L(R)}(E, {\bf k}) = E - \varepsilon ({\bf k}) + 2t_x - (
U^{(0)} + E_c^{(0)} )_{L(R)}$, where, in the last term,
$E_{cL(R)}^{(0)}$ is the bottom of conduction band at the
L(R)-contact and $U^{(0)}_{L(R)}$ is defined by the applied bias.
For instance, if the bias $V$ is applied to the right contact,
then $U^{(0)}_L = 0$ and $U^{(0)}_R = V$.

The advanced GF and the advanced self-energies are Hermitian
adjoints of the corresponding retarded matrices: $G^a = [G^r ]^+$
and $\Sigma^a = [\Sigma^r ]^+$.

In NEGF method the lesser GF plays the central role, in term of
which the measurable quantities are expressed. Noting that in the
device under study the contact coupling is the only 'scattering'
involved, the lesser GF can be expressed as \cite{datta}
\begin{eqnarray}
G^< (E,{\bf k}) & = & G^r (E,{\bf k}) [ \Gamma^<_L (E,{\bf k}) +
              \Gamma^<_R (E,{\bf k}) ] G^a (E,{\bf k}) \nonumber \\
& = & i [ A_L (E,{\bf k}) f_L (E) + A_R (E,{\bf k}) f_R (E)].
\label{7}
\end{eqnarray}
Here, we introduce the lesser tunneling rate defined as
$\Gamma^<_{L(R)} = i[ \Sigma^r_{L(R)} - \Sigma^a_{L(R)} ]
f_{L(R)}(E)$, and the spectral function $A_{L(R)} = G^r [
\Sigma^r_{L(R)} - \Sigma^a_{L(R)}] G^a $. As usual,\cite{potz} the
equilibrium realized in a contact is assumed to be maintained even
in the adjacent region of device. The Fermi function ($f_L$ or
$f_R$) with the same chemical potential ($\mu_L$ or $\mu_R$) can
be then applied in the corresponding region (close to L- or
R-contact).

Once the lesser GF (Eq. \ref{7}) is known it is easy to calculate
the electron density: $[n_e ]_i = -i \int\int dE d{\bf k} [ G^<
]_{i,i}$. Assuming that the vertical effective mass $m_\perp$ is
constant along the $x$-direction, the GF $G^<$ depends on ${\bf
k}$ only through $(E - \varepsilon ({\bf k}))$, the sum over ${\bf
k}$-vector can be then changed into an integral that results in
\begin{equation}
[n_e ]_i = \frac{m_\perp k_B T}{2\pi^2 \hbar^2 a} \int dE
\sum_{a=L,R}[A_a]_{i,i}(E) {\cal F}_0 (\frac{\mu_a - E}{k_B T}); \
i = 1, 2, ..., N, \label{8}
\end{equation}
where ${\cal F}_0$ is the zero-order Fermi-Dirac integral, ${\cal
F}_0 (x) = \ln (1 + e^x )$.

Thus, basically, the main body of the problem is to solve the
Poisson equation (2) and to calculate the electron density (8)
self-consistently. Once the self-consistent solution has been
found, the current can be calculated as
\begin{equation}
I_i \equiv I = I_0 \int dE T(E) \{ {\cal F}_0 (\frac{\mu_L -
E}{k_B T}) - {\cal F}_0 (\frac{\mu_R - E}{k_B T}) \}. \label{9}
\end{equation}
Here $ I_0 = ( e m_\perp k_B T) / (2 \pi^2 \hbar^3 ) $ and the
transmission probability matrix $T(E)$ is defined as
\begin{equation}
T(E) = \Gamma_L (E) G^r (E) \Gamma_R (E) G^a (E), \label{10}
\end{equation}
where $\Gamma_{L(R)} = i ( \Gamma_{L(R)}^r - \Gamma_{L(R)}^a )$ is
the tunneling rate.\cite{datta}

Before presenting the numerical results we would like to mention
that using this self-consistent treatment of the electron
interaction, the NEGF approach used in this work becomes
equivalent to the scattering Fisher-Lee theory. \cite{ferry,datta}
Furthermore, to solve the Poisson equation we use the
Newton-Raphson method with a Jacobian to be determined.
\cite{lake97} In this method it is unnecessary to find the
Jacobian exactly and a good approximation may be acceptable to
reach the convergence. The Jacobian is actually estimated by
fitting the electron density resulting from Eq. (8) with the
electron density expression $[n_e]_i = N_c {\cal F}_{1/2} ((\mu_i
- U_i )/k_B T)$, where $\mu_i$ is the fitting parameter, ${\cal
F}_{1/2}$ is the Fermi-Dirac integral of order $1/2$, and $N_c$ is
the effective conduction band density of states. Using this
procedure, both Eqs. (\ref{1}) and (\ref{2}) are considered as a
single non-linear one for $U$, which really reduces the time of
calculation.

\section{Numerical results}
The numerical calculations have been performed for typical $Al_x
Ga_{1-x}As/GaAs/Al_xGa_{1-x}As$ RTDs. Two barriers are symmetrical
with the height of $0.3 \ eV$ and the thickness of $d = 3 nm$. The
width of quantum well is $w = 5 nm$. The barriers and the quantum
well (scattering region) are undoped. The double-barrier structure
is embedded between two low-doped buffer layers, each $10 nm$ -
thick. The whole system, in turn, is embedded between two outer
access regions of $GaAs$ (emitter and collector), doped at
$10^{18} cm^{-3}$. The thickness of each access region is chosen
as $30 nm$, which is believed to be large enough (see the inset of
Fig. 4). The effective mass is assumed to be constant in the whole
device and equal to $0.067 m_0$, where $m_0$ is the free-electron
mass. The relative dielectric constant of 11.5 is used. This
system exhibits one resonant level associated with the conduction
band minimum, approximately $0.14 eV$ above the bottom of
conduction band of $GaAs$, if the band bending is neglected. The
temperature is generally chosen to be $300 K$. In particular, to
check the device-size and the temperature effects, some
calculations have been also performed for devices with
quantum-well width of $4 nm$, with barrier thickness of $2 nm$
(Figs. 2 and 3), or at temperature of $77 K$ (Figs. 4 and 5).

Here it should be noted that buffer layers with an appropriate
thickness and dopant density are generally used in numerical
calculations as well as in experiments. \cite{ianna98,song,new}
These layers yield a reduction of the highest charge density at
resonance in the quantum well and of the charge accumulation in
the emitter region. Intrinsically such charge accumulations
strongly affect the shape of the potential profile and therefore
produce a current instability. \cite{blant99,potz90} However in
this work we limit our investigations to the structures with
imperceptible instability.

To model the Hamiltonian (3) the 1D-grid of spacing $a = 0.25 nm$
is used. With the algorithm mentioned in the preceding section,
typically, only 5 iterations are required to reach a
self-consistent solution of Eqs. (2) and (8). Fig. $1(a)$ shows
the potential profile calculated for the device with barriers $3
nm$- thick and the quantum-well $5 nm$-wide (hereafter, written
for short as $[3/5/3]$-device) at various applied bias. At zero
bias (solid line), the potential is lightly risen around the
barriers that pulls the resonant level a bit upward. With an
applied bias compared to the resonant one (dashed line) or higher
(dot-dashed line), a large portion of the potential drop occurs at
the right side of the barriers. Qualitatively, our results of
potential profile are similar to those obtained in \cite{potz} for
the $[3/3/3]$-device, using the stationary scattering matrix
approach. A similar potential profile was also recently reported
by Pinaud, \cite{pina} solving self-consistently the
Schr\"{o}dinger and the Poisson equations for the
$[5/5/5]$-device.

The electron density profiles, corresponding to the potentials in
Fig. $1(a)$, are shown in Fig. $1(b)$. Clearly, for any bias, in
the regions where the total potential is zero, the electron
density is precisely given by the donor density. At zero bias the
profile is certainly symmetrical. A finite bias brings about an
electron accumulation in the well. The accumulation reaches the
highest level at the resonant voltage and decreases as the bias is
continuously increased. At the same time, the profile becomes more
and more asymmetrical. For the bias above resonance, a slight
Friedel oscillation of the profile observed in the buffer layer
adjacent to the left barrier is due to an electron accumulation at
the left barrier, which occurs when the resonant level is much
lower than the chemical potential in the left contact. Overall,
these results are again in qualitative agreement with those
presented in, \cite{pina} where however the data for the bias
above resonance is not found.

The most profound manifestation of all that are shown in Figs. 1
can be observed in Fig. $2(a)$, where the transmission coefficient
$T$ is plotted versus the energy $E$ for the same device at
different applied biases. For low bias (less than or about the
resonant bias in I-V characteristics, $V_p \approx 0.31 V$, as can
be seen in Fig. 5), the $T(E)$-curve exhibits sharp peaks of
almost unit on resonance and then falls off rapidly with energy on
both sides. In the range of energy under study we identify two
resonant peaks in each $T(E)$-curve separated by a distance weakly
sensitive to bias and being equal to $\approx 0.24 eV$. The bias,
however, shifts the picture to the left making the first resonant
peak to be cut off when the bias becomes higher than the resonant
one (the case of $ 0.46 V$ in the figure). Such a disappearance of
the sharp peak describes the off-resonant state of the device.
Certainly, both the width of peak as well as the peak-to-peak
distance strongly depend on the device dimensions. In Fig. $2(b)$
we compare the $T(E)$-curves of three devices slightly different
in the barrier thickness or the quantum-well width: $[3/5/3]$
(solid line), $[3/4/3]$ (dashed line), and $[2/5/2]$ (dot-dashed
line). In agreement with the data presented in, \cite{pina} Fig.
$2(b)$ reasonably demonstrates that a decrease of either the
quantum-well width or the barrier thickness makes the resonant
peak wider and the peak-to-peak distance longer. In particular,
two devices with the same quantum-well width of $5 nm$ have the
same position of the first resonant peaks, but the peak width is
larger and the second peak locates at higher energy for the device
with narrower barriers.

In Fig. $3$ we show the I-V characteristics for three devices
corresponding with the transmission coefficients presented in Fig.
$2(b)$. To analyze the data in Fig. $3$ it is convenient to
introduce two quantities: $I_P$ is the current at the peak and
$\gamma = I_P / I_V$ is the peak to valley ratio, where $I_V$ is
the current at the valley. Then, Fig. $3$ demonstrates that a
decrease of either the barrier thickness or the quantum-well width
results in an increase of $I_P$, but a decrease of $\gamma$ (
$\gamma \approx 2.3/0.4$, $\approx 6/1.4$, and $\approx 8/3.4$ for
$[3/5/3]$-(closed circles), $[3/4/3]$-(black squares), and
$[2/5/2]$-(black rhombus)-device, respectively). The latter
statement is a consequence of the smearing of the resonant level
that broadens the peak of $T(E)$ as can be seen in Fig. $2(b)$.
The width of resonance in energy is inversely proportional to the
barrier thickness and/or the quantum-well width. That also
explains, for example, why two devices with the same quantum-well
width of $5 nm$, but with different barrier thicknesses, ($3 nm$
and $2 nm$), exhibit not only strongly different $I_P$, but also
different resonant biases, $V_p \approx 0.307 V$ for
$[3/5/3]$-(closed circles) and $\approx 0.34 V$ for the other,
though the resonant energies are the same, i.e. $\approx 0.14 eV$
(see Fig. $2(b)$).

Concerning the temperature effect we show in Fig. $4$ the I-V
characteristics for the same $[3/5/3]$-device, but at different
temperatures, $300 K $ (closed circles) and $77 K $(black
squares). As the temperature is lowered from $300 K$ to $77 K$,
the current $I_P$ grows from $\approx 2.28 \times 10^5 A cm^{-2}$
to $\approx 2.62 \times 10^5 A
 cm^{-2}$, and simultaneously, $I_V$ falls from $\approx 0.45 \times 10^5 A cm^{-2}$
to $\approx  0.33\times 10^5 A cm^{-2}$, which finally produces a
large change of the peak-to-valley ratio $\gamma$ (from $\gamma
\approx 5$ to $\approx 8$).

As an important note, in the inset of Fig. 4 we compare the I-V
characteristics of two devices with the same quantum-well width
and the same barrier thickness, but with different thicknesses of
the $GaAs$ access region, $30 nm$ (closed circles) and $40 nm$
(crosses). It is apparent that two curves are well coincident in
the whole range of bias under study. This makes an argument to
suggest that, as mentioned in the first section, the $GaAs$ access
region of $30 nm$ thick can be considered as large enough and
consistent with boundary conditions. Such an access region was
used in calculations throughout this work.

\section{Noise calculations}
In this section we demonstrate how the shot noise can be
calculated using the NEGF-code developed in section II. Deviations
of the noise from the full (Poissonian) noise value in RTD have
been extensively investigated in a great number of works, both
theoretical \cite{ianna98,blant99,oriols} and experimental.
\cite{kuzn,song,tsu,nau,new} Mathematically, the measure of these
deviations is the Fano factor $F$ defined as the ratio of the
actual noise spectral density to the full shot noise value $2 e
I$, where $I$ is the average current. Physically, it is widely
accepted that the Pauli exclusion and the charge interaction are
the two correlations, which cause observed shot noise deviations.
While the Pauli exclusion always causes a suppression of noise,
the charge correlation may suppress or enhance the noise,
depending on the conduction regime. In RTD, is was experimentally
found that the noise is partially suppressed (sub-Poissonian
noise, $F < 1$) at low bias voltages (pre-resonance) and becomes
very large (super-Poissonian noise, $F
> 1$) in the negative differential conductance (NDC) region. From theoretical point of view,
typically, there are two approaches based on different
descriptions of tunneling process.

Treating the tunneling through RTD as a quantum coherent process,
a general expression of noise has been derived.
\cite{martin,blant99} In the limit of zero frequency and for
two-terminal structures it has the form:
\begin{eqnarray}
S & = & \frac{e^2}{\pi\hbar} \int dE \int d{\bf k} \{
    T(E_{\bf k})\sum_{\alpha = L,R}f_\alpha (E)(1 - f_\alpha (E))\nonumber \\
  &   &
  + T(E_{\bf k})(1 - T(E_{\bf k})) [f_L (E) - f_R (E)]^2\},\label{14}
\end{eqnarray}
where the first term, proportional to $f_\alpha (E)(1 - f_\alpha
(E))$, describes the thermal noise, and the second term,
proportional to $T(E)[1 - T(E)]$, is nothing but the partition
noise. This expression leads to an important consequence that the
Fano factor should never be less than 1/2. With $T(E)$ determined
in section II, the noise (11) can be straightforwardly calculated.
In the inset of Fig.5 the Fano factor $F$ calculated in this way
for the $[3/5/3]$-device is presented in the range of low
(pre-resonant) bias $V$. It is clear that the obtained
$F(V)$-dependence describes quite well existent experimental data,
\cite{ianna98,song} including a noise suppression in the positive
differential conductance (PDC) region close to the resonance. At
higher bias, in the NDC region, as critically discussed in
\onlinecite{blant99}, the charge interaction becomes to
dramatically affect transport properties of the quantum well, and
the expression (11) can no longer be used.

On the other hand, treating the tunneling as sequential process in
the spirit of master equation, Iannaconne {\it et al.}
\cite{ianna97,ianna98} arrived at the following expression formula
for Fano factor:
\begin{equation}
F = 1 - 2 \tau_g \tau_r /(\tau_g + \tau_r )^2 ,\label{11}
\end{equation}
where the generation and recombination rates through barriers are
given by
\begin{equation}
\tau_g^{-1} = - (dI_L^{(+)} /dN)\mid_{N^*},  \ \  \tau_r^{-1} =
(dI_R^{(-)} /dN)\mid_{N^*} ,\label{12}
\end{equation}
$N$ is the number of electrons in the well at given bias point and
$N^*$ is the steady-state value of $N$, defined from the charge
conservation $I_L^{(+)}(N^*)=I_R^{(-)}(N^*)$. The quantities
$I_L^{(+)}$ and $I_R^{(-)}$ are here the flux of electrons
injected into the well through the left barrier and the flux of
electrons leaving the well through the right barrier,
respectively. It was reported that \cite{ianna98} the Fano factor
calculated from (12) is in good agreement with the experimental
data for the $[12.4/6.2/14.1]$-sample in a large range of bias,
including both the PDC pre-resonant region with a sub-poissonian
noise and the NDC region with a super-poissonian noise. To realize
a calculation of noise (12), using the NEGF-algorithm developed in
section II, we note that the fluxes $I_L^{(+)}$ and $I_R^{(-)}$
can be actually expressed as
\begin{eqnarray}
I_L^{(+)} & = & \int dE \ \Gamma_L (E) [ A_L (E) + A_R (E) ]
{\cal F}_0 (\frac{\mu_L - E}{k_B T}) , \nonumber \\
I_R^{(+)} & = & \int dE \ \Gamma_R (E) [ A_L (E) {\cal F}_0
(\frac{\mu_L - E}{ k_B T}) + A_R (E) {\cal F}_0 (\frac{\mu_R -
E}{k_B T}) ].,
\end{eqnarray}
where $A_{L(R)}$ and $\Gamma_{(L(R)}$ are already defined in
eq.(7) and eq.(10), respectively. So, actually, we can also
calculate the Fano factor $F$ (12), using the NEGF-code developed.
In Fig.5 we show $F$ calculated in this way as a function of bias
for the $[3/5/3]$-device at two temperatures, $77 K$ (closed
circles) and $300 K$ (black squares). Overall, the obtained
$F(V)$-curves qualitatively resemble experimental data \cite{song,
ianna98} with a minimum at the resonant bias and a
super-poissonian peak in the NDC region. For a given device, with
decreasing temperature, as can be seen in Fig. 5, the minimum
becomes deeper (still greater than 1/2) and the super-poissoinian
peak becomes higher. The dramatic increase of noise observed in
experiment \cite{ianna98,song} is caused by the charge interaction
enhanced by the particular shape of the density of state in the
well. In deriving the expression (12) such an interaction has been
self-consistently included. The role of self-consistent Coulomb
interaction in creating a super-Poissonian noise is also recently
demonstrated by Oriols {\it et al.}. \cite{oriols}

Thus, using the NEGF-algorithm developed in section II we are able
to calculate the current noise in both tunneling regimes, coherent
and sequential. It should be here mentioned that in literature
there are different opinions about how a super-Poissonian noise
can be produced and its relation to the NDC. In the theory
\cite{ianna98} the super-poissonian noise is closely associated
with the the NDC, when the time $\tau_g$ (12) is negative. Blanter
and Buttiker emphasized the role of current instability and
stressed that the bias range of super-poissonian noise is
generally different with the NDC range. \cite{blant99} Comparing
the I-V curves and the noises measured in a super-lattice diode
and in a RTD, Song {\it et al.}\cite{song} concluded that not NDC,
but charge accumulation in the well responds for the
super-poissonian noise observed in RTD. Safanov {\it et al.}
\cite{safanov}, measuring the noise in resonant tunneling via
interacting localized states, observed a super-Poissonian noise in
the range of bias, where there is no NDC. Authors have also
pointed out that the effect on noise of the Pauli exclusion
principle and of the Coulomb interaction are similar in most
mesoscopic systems. For Coulomb blockade metallic quantum dot
structures, studies \cite{apl,prb} support the idea that not NDC,
but charge accumulation in dots, responds for super-poissonian
noise. Remarkably, Aleshkin {\it et al.}\cite{alesh} recently
claimed that while in the sequential tunneling double barrier
model the Fano factor $F$ is still limited by the lowest value of
$1/2$, in the coherent model $F$ may drop below this value. Then,
authors also suggest that the noise suppression below the value
$1/2$ can be used to identify a coherent transport. Actually, to
solve all these, very fundamental, contradictions a more
systematic analysis of noise, taking adequately into account the
charge interaction is requested. To this end, it is perhaps most
convenient to use the NEGF, well-known as the quantum transport
approach capable even to deal with far from equilibrium systems.

\section{Conclusion} The NEGF approach has been formulated and
implemented in a fully
self-consistent calculating procedure that can be readily applied
to any 1D nanoscale structures and extended to more complicated
devices, e.g. nanoscale field effect transistors. The Poisson
equation solver routine has been improved to considerably speed up
calculations. It yields physically reasonable results and allows
us to work in a large range of temperature.

Numerical calculations have been performed for RTDs of different
designs and at different temperatures. The potential and electron
density profiles obtained for various applied bias, below and
above resonance, rationally describe the resonant behavior in the
device. The transmission coefficient and the corresponding I-V
characteristics seem to be sensitive not only to the quantum-well
width, but also to the barrier thickness. Besides the well-known
fact that the quantum-well width defines the resonant level, our
calculations show that a decrease of either the quantum-well width
or the barrier thickness makes the peak of transmission
coefficient wider and the peak-to-peak distance longer.
Correspondingly, the peak-to-valley ratio in I-V curves decreases,
though the value of the current at the peak is considerably risen.
The reason merely lies in a broadening of the resonant level,
caused by narrowing the quantum-well width and/or the barrier
thickness. Additionally, by lowering the temperature from $300 K$
to $77 K$ we observed not only a raise of the peak in I-V curve,
but also a reduction of the valley current that leads to a
significant increase in the peak-to-valley ratio.

It was also shown that the NEGF-algorithm developed can be
straightforwardly used to calculate the current noise in both
coherent and sequential tunneling models. In qualitative agreement
with experiments, obtained results highlight the role of charge
interaction which causes a fluctuation of density of states in the
well and therefore a noise enhancement in the NDC region. We
believe that in this way, i.e. using the NEGF, a fully quantum
transport approach for non-equilibrium problems, one can calculate
the noise in a variety of interacting mesoscopic systems and
therefore better understand the nature of noise deviations as well
as of conduction mechanisms in these structures.

{\bf Acknowledgments}. This work has been partially done with the
support of the European Community under contract IST-506844 ($No$.
E SINANO). One author (V.L.N) thanks the CNRS for partial
financial support under PICS programme $No$. 404 to his visit at
IEF where this work has been done.

\newpage
\begin{figure}[htp]
\begin{center}
\caption{\label{fig:1} Potential profile (a) for the
$[3/5/3]$-device and corresponding charge density (b) at various
applied bias: $0 V$ (solid line), $0.31 V$ (close to resonance,
dashed line), and $0.46 V$ (above resonance, dot-dashed line),
Temperature: $300 K$.}

\caption{\label{fig:2} (a) Transmission coefficient in logarithmic
scale for the $[3/5/3]$-device at $300 K$ at the applies bias:
$0.31 V$ (dashed line), and $0.46 V$ (dot-dashed line). (b)
Comparison the transmission peak of different structures: [3/5/3]
(solid line), [3/4/3] (dashed line), and [2/5/2] (dot-dashed line)
at V = 0.00V.}

\caption{\label{fig:3} I-V characteristics for the same devices as
in Fig.4: $[3/5/3]$-(closed circles), $[3/4/3]$-(black squares),
and $[2/5/2]$-(black rhombus); Temperature: $300 K$.}

\caption{\label{fig:4} The temperature effect: $300 K$(closed
circles) and $77 K$ (black squares) for the same $[3/5/3]$-device.
Inset: the I-V characteristics for two devices with the same
quantum-well width and the same barrier thickness, $[3/5/3]$, but
with different thicknesses of the $GaAs$ access region, $30 nm
(\bullet)$ and $40 nm (\times)$, are compared. Temperature: $300
K$.}

\caption{\label{fig:5} The Fano factor calculated from Eq.(13) is
plotted as a function of the applied bias for the $[3/5/3]$-device
at $300 K $ (black square) and $77 K $ (closed circles). Inset:
the Fano factor calculated from B\"{u}ttiker's formula (11) for
the same device at $77 K$.}
\end{center}
\end{figure}

\newpage
\vspace*{5cm}
\begin{figure}[htp]
\begin{center}
\includegraphics{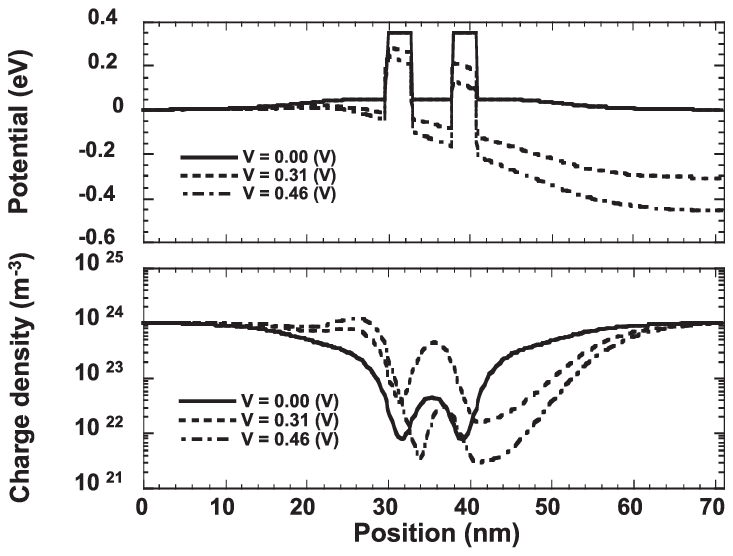}
\end{center}
\end{figure}
\vspace{3cm} FIGURE 1 (V. Nam DO)

\newpage
\vspace*{5cm}
\begin{figure}[htp]
\begin{center}
\includegraphics{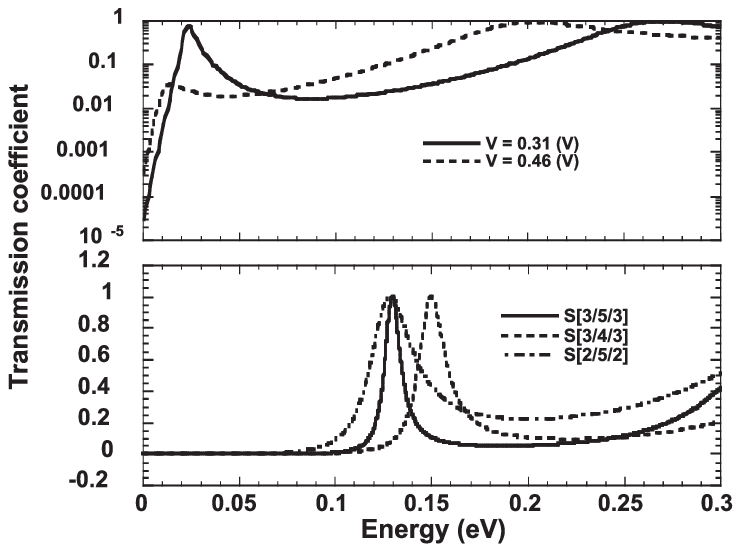}
\end{center}
\end{figure}
\vspace{3cm} FIGURE 2 (V. Nam DO)

\newpage
\vspace*{5cm}
\begin{figure}[htp]
\begin{center}
\includegraphics{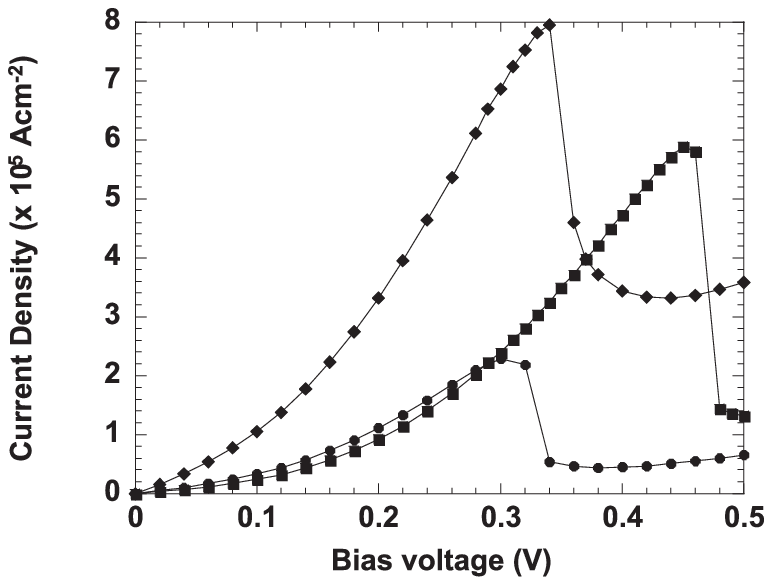}
\end{center}
\end{figure}
\vspace{3cm} FIGURE 3 (V. Nam DO)

\newpage
\vspace*{5cm}
\begin{figure}[htp]
\begin{center}
\includegraphics{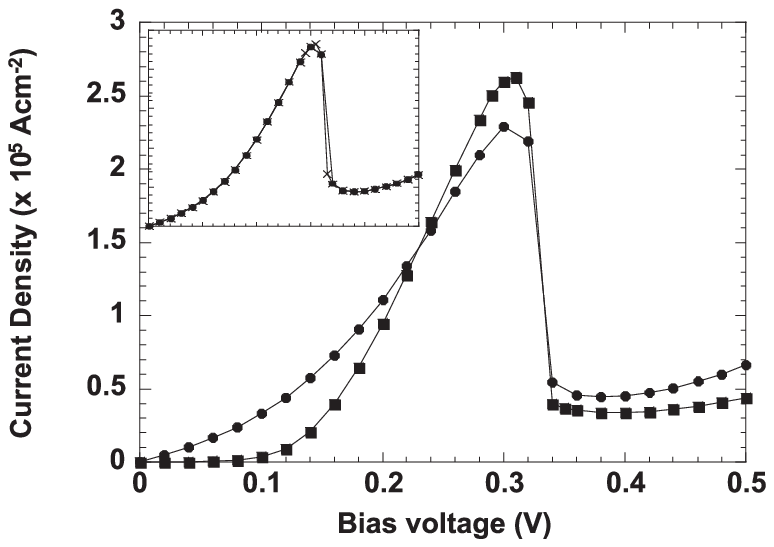}
\end{center}
\end{figure}
\vspace{3cm} FIGURE 4 (V. Nam DO)

\newpage
\vspace*{5cm}
\begin{figure}[htp]
\begin{center}
\includegraphics{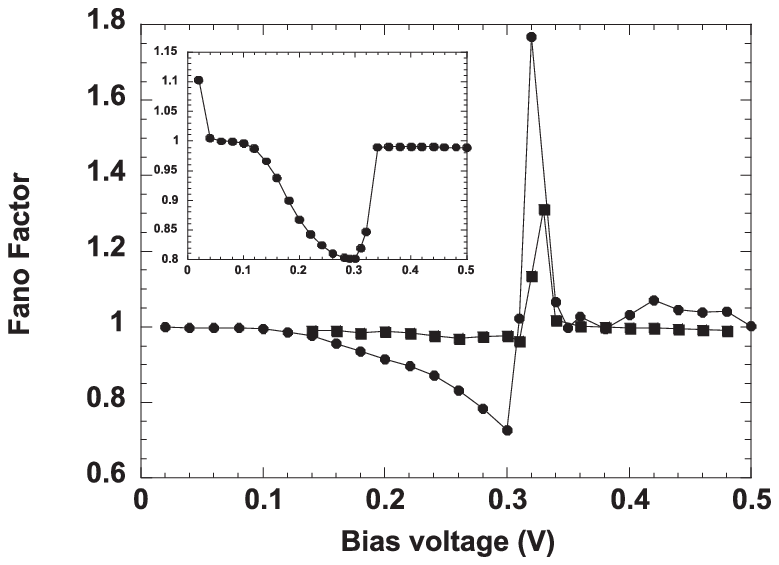}
\end{center}
\end{figure}
\vspace{3cm} FIGURE 5 (V. Nam DO)

\end{document}